\documentclass[10pt,conference]{IEEEtran}
\IEEEoverridecommandlockouts
\usepackage{cite}
\usepackage{amsmath,amssymb,amsfonts}
\usepackage{algorithm}
\usepackage{algpseudocode}
\usepackage{graphicx}
\usepackage{textcomp}
\usepackage{xcolor}
\def\BibTeX{{\rm B\kern-.05em{\sc i\kern-.025em b}\kern-.08em
    T\kern-.1667em\lower.7ex\hbox{E}\kern-.125emX}}

\makeatletter
\def\ps@IEEEtitlepagestyle{%
\def\@oddfoot{\mycopyrightnotice}%
\def\@evenfoot{}%
}
\def\mycopyrightnotice{%
{\footnotesize © 2025 IEEE. Accepted for publication in 2025 IEEE Globecom Workshops (GC Wkshps), Taipei, Taiwan.\hfill} 
\gdef\mycopyrightnotice{}
}

\begin{document}

\title{End-to-end Learning of Probabilistic and Geometric Constellation Shaping with Iterative Receivers\thanks{This work is a part of the Finnish Ministry of Education and Culture’s Doctoral Education Pilot under Decision No. VN/3137/2024-OKM-6 (The Finnish Doctoral Program Network in Artificial Intelligence, AI-DOC). A part of this work is supported by the 6GARROW project which has received funding from the SNS-JU under the EU’s Horizon Europe research and innovation programme (Grant No. 101192194) and from the IITP grant funded by the Korean government (MSIT) (Grant No. RS-2024-00435652).}}



\author{\IEEEauthorblockN{Harindu Jayarathne,
Dileepa Marasinghe,
Nandana Rajatheva,
and
Matti Latva-aho}
\IEEEauthorblockA{\textit{Center for Wireless Communications} \\
\textit{University of Oulu, Finland}\\
E-mail: harindu.ravin@oulu.fi, dileepa.marasinghe@oulu.fi, nandana.rajatheva@oulu.fi, matti.latva-aho@oulu.fi}
}

\maketitle

\begin{abstract}
An end-to-end learning method for constellation shaping with a shaping-encoder assisted transceiver architecture is presented. The shaping encoder, which produces shaping bits with a higher probability of zeros, is used to produce an efficient symbol probability distribution. Both the probability distribution and the constellation geometry are jointly optimized, using end-to-end learning. Optimized constellations are evaluated using two iterative receiver architectures. Bit error rate (BER) performance gain is quantified against standard amplitude phase-shift keying (APSK) and quadrature amplitude modulation (QAM) constellations. A maximum BER gain of 0.3 dB and 0.15 dB are observed under two receivers for the learned constellations compared to standard APSK or QAM. The basic approach is extended to incorporate the full iterative detection and decoding loop, using the deep unfolding technique. A bit error rate gain of 0.1 dB is observed for the iterative scheme with learned constellations under block fading channel conditions, when compared to standard APSK. 
\end{abstract}


\section{Introduction}

Constellation shaping maximizes the transmit information rate by optimizing the transmit signal distribution. This can be achieved using geometric shaping, probabilistic shaping, or a combination of both, where signal point positions and their probabilities are jointly optimized. In probabilistic shaping, the low energy symbols are transmitted more frequently than high energy symbols. Effectively, it minimizes the average transmit energy required to transmit information at a certain rate with an acceptable bit error rate. Maximizing the mutual information $I(X;Y)$ with respect to the constellation geometry or the input probability distribution is a well-known approach \cite{huang2005characterization,ozaydin2022grand, stark2019joint} for constellation shaping. 


To practically implement a non-uniform input probability distribution in bit-interleaved coded modulation (BICM) \cite{caire1998bicm} systems, a mechanism is needed for mapping uniformly distributed information bits to non-uniformly distributed constellation symbols. Probabilistic amplitude shaping (PAS) \cite{bocherer2015bandwidth} scheme combined with constant composition distribution matching \cite{schulte2015constant} is a well-known pragmatic approach with minimum compromise in receiver complexity. Despite few restrictions in effective information rate and constellation geometry, the novel placement of the forward error correction (FEC) encoder after the distribution matcher avoids accumulation of errors in the distribution dematcher. Different from PAS architecture, the constellation shaping scheme proposed in \cite{le2005bit, le2007constellation, valenti2012constellation} uses a shaping encoder as an alternative to the distribution matcher. The shaping encoder takes in uniformly distributed bits and produces a reversible non-uniform bit-stream for selected bit-inputs to the symbol mapper. Thereby, the symbol mapper produces low energy symbols with high probability. Since the shaping encoder is placed after the FEC encoder, an iterative receiver architecture is required to achieve a useful bit error rate (BER) gain. Within this scheme, authors of \cite{valenti2012constellation} consider standardized APSK constellations and optimize the probability distribution and the constellation geometry using a grid search approach. Furthermore, the authors of \cite{le2005bit, le2007constellation} consider a fixed pulse-amplitude modulation constellation geometry, which simplifies the probability distribution optimization.

\begin{figure*}[htbp]
\centerline{\includegraphics[width=0.8\textwidth]{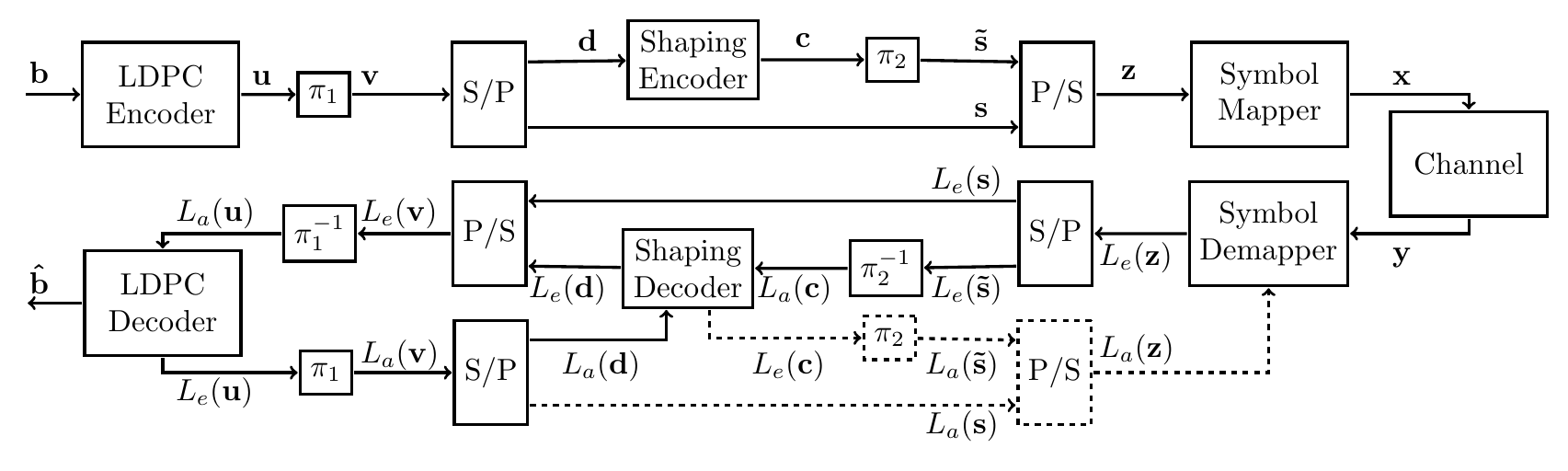}}
\caption{End-to-end architecture of the constellation shaping scheme.}
\label{PSsystemfig}
\end{figure*}

In end-to-end learning for wireless communication systems, the transmitter and the receiver are jointly trained as a single neural network across a differentiable channel\cite{trainable_comm}. Recent studies have shown that end-to-end learning is an excellent approach for constellation shaping \cite{stark2019joint,ait2020joint,trainable_comm, marasinghe2024constellation}. Joint learning of geometric and probabilistic constellation shaping is initially studied in \cite{stark2019joint}. Authors of \cite{ait2020joint} generalize the PAS architecture and investigate joint learning of geometric and probabilistic constellation shaping for coded modulation systems. Authors of \cite{rode2023end} explore this approach in the context of optical communications when Wiener phase noise is present and carrier phase estimation is performed. However, the end-to-end learning design is tailored to PAS architecture, where the FEC encoder is placed after the distribution matcher and uniformly distributed parity bits are directly sent as input to the symbol mapper. End-to-end learning for a probabilistic constellation shaping system, where the distribution matching function is placed after the FEC encoder remains to be explored. In this research, joint learning of geometric and probabilistic constellation shaping is investigated, extending the shaping encoder-assisted constellation shaping scheme detailed in \cite{valenti2012constellation}. Two suitable training architectures are investigated for the iterative detection and decoding (IDD) and non-IDD receiver architectures considered in the paper. The former includes the FEC decoder and the shaping decoder blocks within the training architecture, which is designed using the deep unfolding technique. BER performance is evaluated under additive white Gaussian noise (AWGN) and block fading channel conditions. The contributions of this paper are the following:

\begin{enumerate}
    \item We investigate an end-to-end learning method for joint geometric and probabilistic constellation shaping for a shaping-encoder assisted constellation shaping scheme. Two training architectures are proposed: one for the IDD receiver and another for its non-IDD low-complexity variant. The former incorporates the iterative detection and decoding loop within the training architecture. 
    \item The resulting performance under AWGN and block fading channels is evaluated. Constellations trained under the non-IDD training architecture exhibit a maximum of 0.15 dB and 0.3 dB BER gain for two receivers compared to when standard constellations are used. A BER gain reaching 0.1dB is observed under block fading channel with perfect channel state information (CSI) when the IDD training architecture is used.

\end{enumerate}

\section{System Model}

\subsection{Transmitter Design}

Consider the transmitter architecture shown in Fig. \ref{PSsystemfig}. The architecture is akin to the well-known BICM transmitter and additionally includes a shaping encoder block for generating a non-uniform symbol probability distribution. Firstly, uniformly distributed information bit stream $\mathbf{b}$ of length $k$ is encoded by the low density parity check (LDPC) FEC encoder of rate $R_c = \frac{k}{n}$ to obtain $\mathbf{u}$ of length $n$. The encoded bit stream is subsequently permuted using a random bit interleaver $\pi_1$ to obtain $\mathbf{v}$. The shaping encoder maps input bit sequences of length $k_s$ with output bit sequences of length $n_s$, operating at a rate of $R_s = \frac{k_s}{n_s}$. A segment of length $Lk_s$ is extracted from $\mathbf{v}$ to obtain $\mathbf{d}$, where $L$ is a positive integer. The bit sequence $\mathbf{d}$ is encoded by the shaping encoder to obtain $\mathbf{c}$ of length $Ln_s$. The encoder is structured to generate output sequences that contain a high density of zeros with a probability of $p_0 > \frac{1}{2}$ \cite{calderbank1990nonequiprobable, le2005bit, valenti2012constellation}. After permuting $\mathbf{c}$ using a second bit interleaver $\pi_2$, $\mathbf{\tilde{s}}$ is obtained. Rest of the bits in $\mathbf{v}$ are arranged into $\mathbf{s}$. Both $\mathbf{s}$ and $\mathbf{\tilde{s}}$ are combined to form $\mathbf{z}$, which is modulated by the symbol mapper using a complex constellation $\mathcal{X}$. Modulation order of $\mathcal{X}$ is $m = \log_2(M)$ bits/symbol, where $M$ is the constellation size. Among the $\{1,\hdots,m\}$ bit inputs to the mapper, the subset of shaped bit indices is given by $\mathcal{S}$, where $0 < |\mathcal{S}| < m$ holds. The subset of unshaped bit indices are given by $\mathcal{\tilde{S}}$. Symbols having labels with zeros in the selected $\mathcal{S}$ bit positions have a higher probability of occurrence. The aim is to divide the constellation into $2^{|\mathcal{S}|}$ sub-constellations and assign sub-constellation specific symbol probabilities. It follows that $|S| = \frac{mLn_s}{n-Lk_s +Ln_s}$ holds and is assumed that the system parameters are selected such that $L$ is an integer. We denote the binary random variable attributable to the $k$\textsuperscript{th} bit input to the mapper as $B_k$. With $P(B_k = 0) = p_0$ for $k \in \mathcal{S}$ and $P(B_k = 0) = 1/2$ for $k \in \mathcal{\tilde{S}}$, the symbol probability distribution $P_{X}(x)$ is computed using
\begin{equation}
   P_X(x) = \prod_{k =1}^{m} P(B_k = f_k(x)),
\label{bittosymbprob}
\end{equation}
where the $k$\textsuperscript{th} bit of the bit-label associated with the symbol $x$ is given by $f_k(x)$. By carefully selecting the bit-labeling and the geometry of the constellation, low energy symbols can be assigned higher probabilities to obtain an efficient constellation.

The modulated transmit symbol stream $\mathbf{x}$ is sent through the channel to obtain the received symbols $\mathbf{y}$. Considering the $j^{\text{th}}$ element of $\mathbf{x}$, the channel input $x_j$ and output $y_j$ is related by $y_j = h_j x_j + n_j$, where $n_j$ denotes zero mean complex Gaussian noise of variance $N_0$, and $h_j$ denotes the fading channel coefficient. We assume that $h_j$ is constant across a block of $N_{\text{BF}}$ symbols, with $N_p$ pilot symbols inserted per block to facilitate channel estimation. The AWGN channel is realized when $h_j = 1$ constant over all symbols. The overall information rate of the shaped system is given by,
\begin{equation}
R = R_c(m + |\mathcal{S}|(R_s-1)).
\end{equation}
To compare the BER performance of the shaped system with a uniform system, an FEC encoder rate of $R/m$ is required in the uniform system.

\subsection{Receiver Design}

The receiver architectures for the shaping encoded transmitter is well investigated in \cite{le2007constellation, valenti2012constellation}. Two types of receivers are considered in this paper: 1) an iterative detection and decoding (IDD) receiver and 2) a simplified low complexity receiver \cite{le2007constellation}. The IDD receiver is indicated in Fig. \ref{PSsystemfig}. The simplified receiver architecture is obtained when the components shown with dashed lines are excluded. In this receiver, the demapper is excluded from the iterative decoding loop and the soft information is iterated only between the FEC decoder and the shaping decoder. We use the term \lq simplified scheme' to denote the combined transmitter and the simplified receiver in the rest of the paper. 

Firstly, the symbol demapper computes the output extrinsic log-likelihood ratios (LLRs) $L_e(\mathbf{z})$ using the received symbols. For AWGN channel, the $k$\textsuperscript{th} LLRs attributable to the $j^{\text{th}}$ transmit symbol of $\mathbf{x}$ is given by
\begin{equation}
\begin{split}
&L_e(z_j(k)) \\&= \ln \frac{
\displaystyle\sum_{\hat{x}_j \in \mathcal{X}_1^k} \exp \left(
-\frac{\left\lvert y_j-\hat{x}_j\right\rvert^2}{N_0} +
\displaystyle\sum_{\substack{n=1 \\ n \neq k}}^{m} \! f_n(\hat{x}_j) L_a(z_j(n))
\right)
}{
\displaystyle\sum_{\hat{x}_j \in \mathcal{X}_0^k} \exp \left(
-\frac{\left\lvert y_j-\hat{x}_j\right\rvert^2}{N_0} +
\displaystyle\sum_{\substack{n=1 \\ n \neq k}}^{m} \! f_n(\hat{x}_j) L_a(z_j(n))
\right),
} 
\end{split}
\end{equation}
where $k \in \{1,\hdots,m\}$ and $\mathcal{X}_b^k$ denotes the set of symbols in $\mathcal{X}$, whose $k$\textsuperscript{th} bit-label is $b \in \{0,1\}$ \cite{le2007constellation}. For both schemes, the a priori information $L_a(\mathbf{z})$ is initialized with $L_a(\mathbf{s}) = 0$ and $L_a(\mathbf{\tilde{s}}) = \log{\frac{1-p_0}{p_0}}$. After serial-to-parallel and deinterleaver operations, the shaping decoder estimates the LLRs corresponding to the input bits of the shaping encoder $\mathbf{d}$. With the apriori information $L_a(\mathbf{c})$ and $L_a(\mathbf{d})$ available, the shaping decoder computes the soft estimate of the $j^\text{th}$ bit of $\mathbf{d}$ using
\begin{equation}
L_e(d_j) = \ln \frac{
\displaystyle\sum_{\mathbf{\hat{d}} \in \mathcal{D}_1^j} \exp \left(
\displaystyle\sum_{n=1}^{n_s} \! f_n(\mathbf{\hat{d}}) L_a(c_n) +
\displaystyle\sum_{\substack{\ell=1 \\ \ell \neq j}}^{k_s} \! \hat{d}_\ell L_a(d_\ell)
\right)
}{
\displaystyle\sum_{\mathbf{\hat{d}} \in \mathcal{D}_0^j} \exp \left(
\displaystyle\sum_{n=1}^{n_s} \! f_n(\mathbf{\hat{d}}) L_a(c_n) +
\displaystyle\sum_{\substack{\ell=1 \\ \ell \neq j}}^{k_s} \! \hat{d}_\ell L_a(d_\ell)
\right),
}
\label{shapingdeceq}
\end{equation}
where $f_n(\mathbf{\hat{d}})$ denotes the $n^{\text{th}}$ bit of the output bit sequence attributable to the input bit sequence $\mathbf{\hat{d}}$ and $\mathcal{D}_b^j$ denotes the set of bit sequences, whose $j$\textsuperscript{th} bit is $b \in \{0,1\}$. The $\ell$\textsuperscript{th} element of $\mathbf{d}$ and $\mathbf{\hat{d}}$ are denoted by $d_\ell$ and $\hat{d}_\ell$, respectively. The apriori information $L_a(\mathbf{d})$ is initially set to zero. Remaining parallel-to-serial and deinterleaver blocks reverse the operation of corresponding blocks in the transmitter. Given that belief propagation decoder estimates $L_\text{FEC}(\mathbf{u})$ using $L_a(\mathbf{u})$, the extrinsic information at the FEC decoder is computed using
\begin{equation}
    L_e(\mathbf{u}) = L_\text{FEC}(\mathbf{u}) - L_a(\mathbf{u}).
\end{equation}
In the simplified scheme, soft information from the FEC decoder is sent back to the shaping decoder as a priori information to initiate the next iteration between them. After required number of iterations between the decoder and the shaping decoder, a hard estimate $\mathbf{\hat{b}}$ of the message bits is obtained from the decoder.

In the IDD receiver, the shaping decoder recomputes the LLRs corresponding to $\mathbf{c}$ after FEC decoding. With $L_a(\mathbf{c})$ computed before decoding and $L_a(\mathbf{d})$ computed after decoding, the shaping decoder computes the soft estimate of the $j^\text{th}$ bit of $\mathbf{c}$ denoted by $L_e(c_j)$ using a similar approach to \eqref{shapingdeceq} \cite{valenti2012constellation}. After required number of iterations between the decoder and the demapper, a hard estimate $\mathbf{\hat{b}}$ is obtained. Note that the shaping decoder computation is analogous to maximum a posteriori (MAP) demapper computation, and the shaping decoder computation has the differentiable property, enabling gradient-based optimization.



\section{End-to-end System Training}

For simplified and IDD receivers, two different training architectures are proposed for constellation shaping. For both architectures, the end-to-end training is performed between the bit input to the mapper and the demapper LLR output. However, the second architecture incorporates the FEC decoder and shaping decoder within the training loop.

\subsection{Non-IDD Training Architecture}

\begin{figure}[b]
\centerline{\includegraphics[width=\columnwidth]{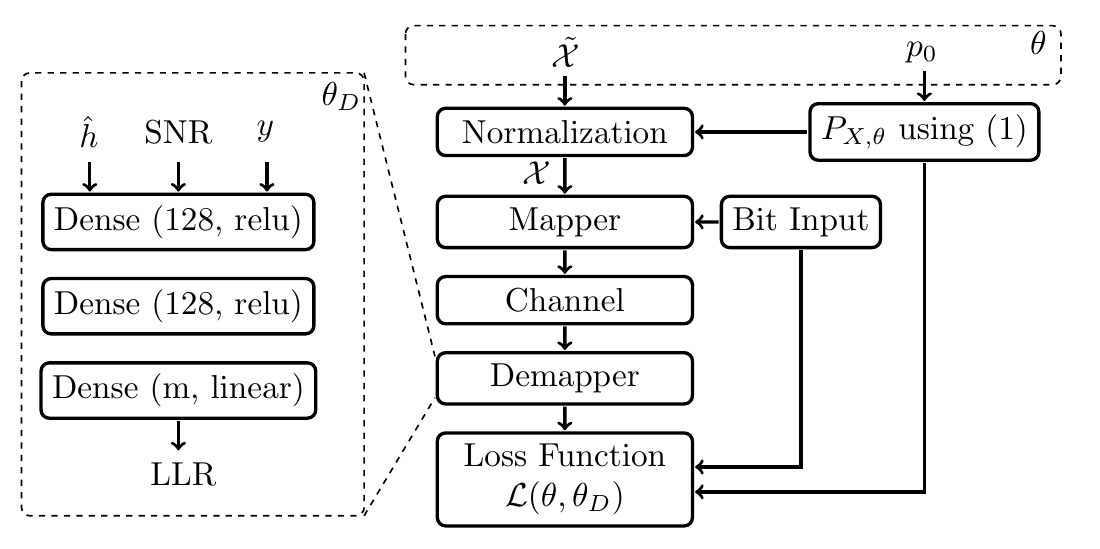}}
\caption{The training setup aimed for the simplified receiver. Training is performed between the mapper input and the demapper output.}
\label{trainingsetup}
\end{figure}

In the simplified scheme, no a priori information is fed back to the demapper from the decoder. Therefore, the constellation geometry $\mathcal{X}$ and $P_{X,\theta}$ are optimized such that information rate resulting from bit metric decoding at the BICM receiver is maximized. Fig. \ref{trainingsetup} shows the end-to-end training setup used for joint probabilistic and geometric constellation shaping. The trainable parameters $\theta$ includes the unnormalized constellation points $\tilde{\mathcal{X}} \in \mathbb{R}^{2M}$ and $p_0 \in [0,1]$. With $p_0$ available, the symbol probability distribution $P_{X,\theta}$ is computed using \eqref{bittosymbprob}, which is subsequently used for unit average energy normalization of the constellation, probability aware demapping and loss function calculation. Normalization of $\tilde{\mathcal{X}}$ for unit average energy yields $\mathcal{X}$. A batch of transmit symbols with equal occurrences of all constellation symbols from $\mathcal{X}$ is sent through the channel. Note that the loss function subsequently incorporates the symbol probabilities by appropriately weighting the terms using $P_{X,\theta}$. A neural demapper with trainable parameters $\theta_D$ is used to obtain the LLR estimates. With the channel estimate, equalized received signal and $E_b/N_0$ value as input, the neural demapper consists of two dense layers with relu activation function and a final dense layer with linear activation function. For AWGN channel, the optimal MAP demapper is used instead of the neural demapper, since the differentiable MAP demapper function allows gradients to flow during training. Note that the probability of a symbol point depends both on its bit-labeling and $p_0$, while each symbol point has the freedom to move on the in-phase and quadrature plane during training.


Following the approach proposed in \cite{ait2020joint}, the information rate is approximated with the loss function used for training. The LLRs are used to compute the loss function 
\begin{equation}
    \mathcal{L}(\mathbf{\theta},\mathbf{\theta}_D) = \mathcal{L}_{\text{BCE}}(\mathbf{\theta},\mathbf{\theta}_D) - H_{P_{X,\theta}}, 
    \label{overallloss}
\end{equation}
where $H_{P_{X,\theta}} = -\sum_{x \in \mathcal{X}}  P_{X,\theta}(x)\log(P_{X,\theta}(x))$ is the entropy of the random variable $X$ representing the channel input and $\mathcal{L}_{\text{BCE}}(\mathbf{\theta},\mathbf{\theta}_D)$ denotes the binary cross entropy (BCE) loss between the input bit values to the mapper and the demapper output probabilities. The estimated posterior distribution $\tilde{P}_{\theta_D}(b|y)$, for $b \in \{0,1\}$, is obtained by applying the Sigmoid function to the demapper output LLRs, where $y$ denotes a random channel output. The value of $\mathcal{L}_{\text{BCE}}(\mathbf{\theta},\mathbf{\theta}_D)$ is computed by
\vspace{-2mm}
\begin{equation}
\begin{aligned}
& \mathcal{L}_{\text{BCE}}(\mathbf{\theta},\mathbf{\theta}_D)  =  \sum_{k=1}^{m} \underset{b_k, y}{\mathbb{E}} [ - \log(\tilde{P}_{\theta_D}(b_k|y))] \\& = - \frac{1}{B}\! \sum_{j=1}^{B}\sum_{k=1}^{m} \sum_{x^j \in \mathcal{X}} \sum_{b^j_k = 0}^1 \! \! P(b^j_k|x^j)P_{X,\theta}(x^j)\log(\tilde{P}_{\theta_D}(b^j_k|y^j)),
\end{aligned}
\label{bceloss}
\end{equation}
where $B$ denotes the batch size and $P(b_k|x) \in \{0,1\}$ represents whether $f_k(x) = b_k$ holds for $b_k \in \{0,1\}$. 


\subsection{IDD Training Architecture}

\begin{figure}[t]
\centerline{\includegraphics[width=\columnwidth]{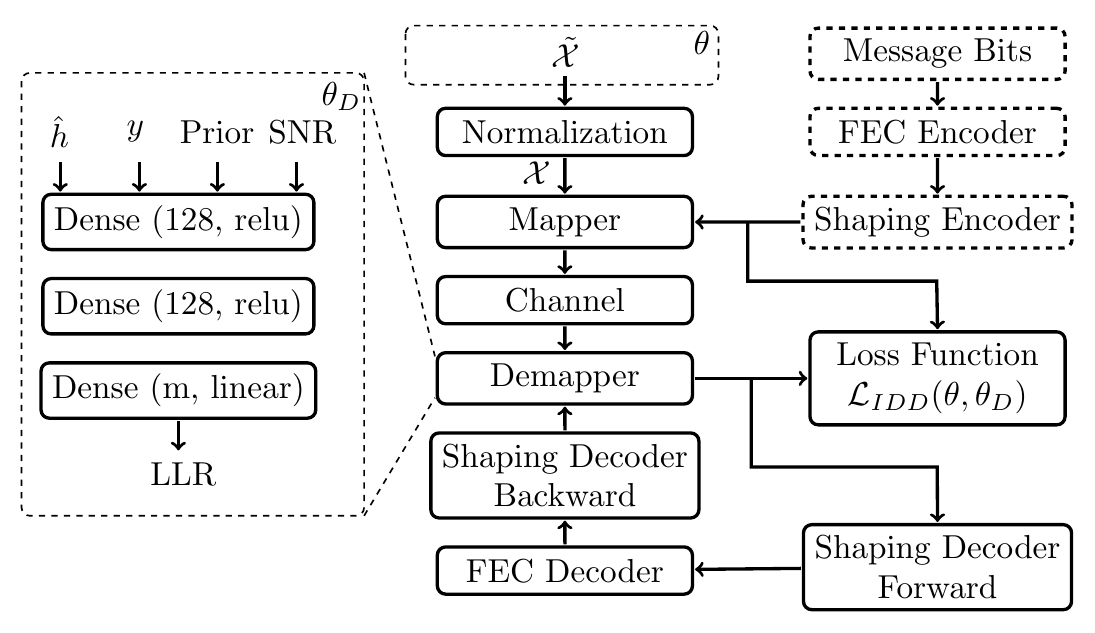}}
\caption{The training setup aimed for the IDD receiver. Training is performed between the mapper input and the demapper output, with both the FEC decoder and the shaping decoder integrated into the model architecture.}
\label{trainingsetup_Idd}
\end{figure}

In this section, the previous approach is extended to incorporate the shaping decoder and FEC decoder into the training loop. A predetermined shaping encoder is used and corresponding symbol probabilities are considered as constant. As shown in Fig. \ref{trainingsetup_Idd}, actual codewords are used to train the IDD system using the deep unfolding technique \cite{trainable_comm}. Since the operations in both the belief propagation decoder and shaping decoder in Fig. \ref{PSsystemfig} satisfy the differentiable property, the gradients can be back-propagated through the iterative algorithm, enabling end-to-end training. Different from the conventional IDD systems, a single iteration of belief propagation (BP) decoding is implemented within each IDD iteration for both training and evaluation. We use the loss function proposed in \cite{trainable_comm} for BICM-ID systems given by
\begin{equation}
\begin{aligned}
\mathcal{L}_{\text{IDD}}(\mathbf{\theta},\mathbf{\theta}_D)  =  \sum_{i=1}^{I}\sum_{k=1}^{m} \underset{b_k, y}{\mathbb{E}} [ - \log(\tilde{P}_{\theta_D,i}(b_k|y))],
\end{aligned}
\label{bcelossidd}
\end{equation}
where $I$ denotes the total number of iterations between the demapper and the FEC decoder. The binary cross entropy loss values at the output of the demapper across all IDD iterations are accumulated by $\mathcal{L}_{\text{IDD}}(\mathbf{\theta},\mathbf{\theta}_D)$. Note that the entropy term, which appears in \eqref{overallloss}, is omitted in \eqref{bcelossidd}, as the symbol probabilities are fixed.

\section{Results}

First, the outlined non-IDD training architecture is used to obtain learned constellations and their performance is evaluated under AWGN channel. Next, the results are extended to evaluate the deep unfolding based IDD training architecture under both the AWGN and block fading channel conditions. Training and evaluation are conducted using a single MIG-1G.5GB instance of an NVIDIA A100-SXM4-40GB graphics processing unit.

\subsection{Non-IDD Training Architecture}
\label{subsec:non_idd}

To evaluate the performance of the learned constellations, we compare the performance under two baseline constellations: 32-APSK and 32-QAM. Since the shaping encoder output is fed into the $S$ bit positions, the bit-labeling of both standard 32-APSK and Gray-labeled 32-QAM are modified to have zero in the $S$ bit positions of the low energy symbols. Bit labels are flipped or swapped across selected bit positions while preserving the hamming distances between symbol labels \cite{valenti2012constellation}. Training is performed using the Adam \cite{kingma2015adam} optimizer with a learning rate $10^{-3}$, a batch size $B = 10^3$, and 5000 training iterations. Within the training batch, AWGN noise samples are generated such that associated $E_b/N_0$ value of each batch item is evenly spread between 5 dB and 6 dB. 
\begin{figure}[!t]
\centerline{\includegraphics[width=0.9\columnwidth]{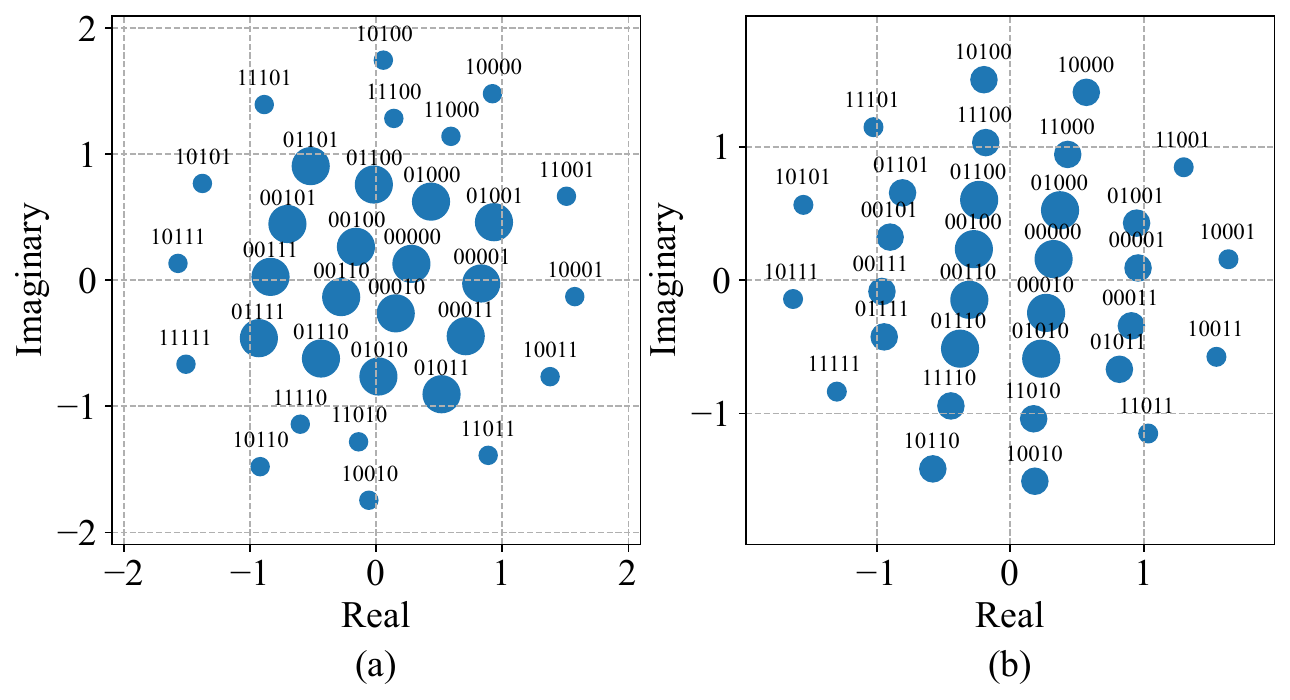}}
\caption{(a) Learned constellation $T_1$ obtained under the AWGN channel for $S = \{0\}$ system. (b) Learned constellation $T_2$ obtained under AWGN channel for $S = \{0,4\}$ system. The diameter of the symbol points indicates their respective probability of occurrence.}
\label{learnedconst}
\end{figure}
\begin{figure}[!t]
\centerline{\includegraphics[width=0.9\columnwidth]{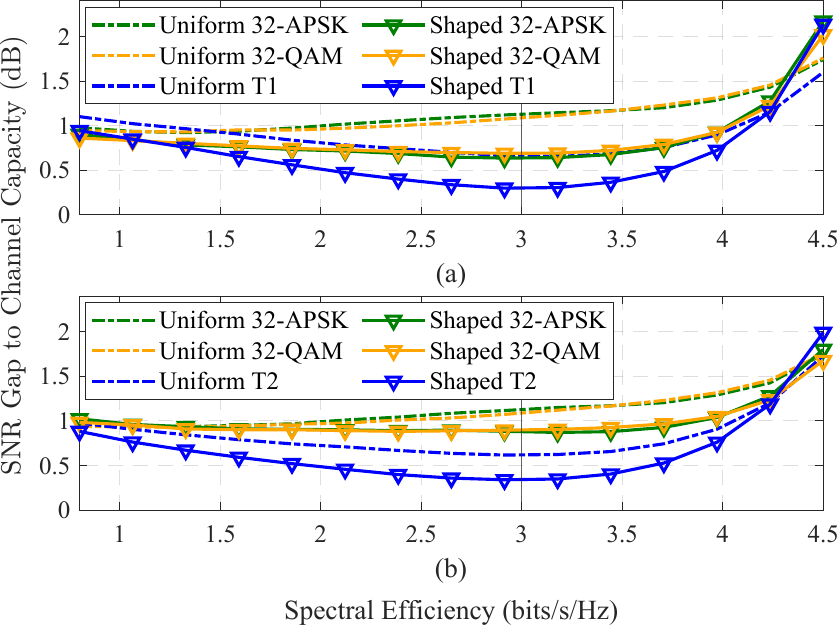}}
\caption{Signal-to-noise ratio (SNR) gap to Gaussian channel capacity of BICM capacity for uniform and shaped systems with 32-QAM, 32-APSK and trained constellations.}
\label{gaptocap}
\end{figure}
Fig. \ref{learnedconst}(a) and Fig. \ref{learnedconst}(b) contain two learned constellations obtained for $S = \{0\}$ and $S = \{0,4\}$ cases, which we refer to as $T_1$ and $T_2$. For both cases, $P_{X,\theta}$ converged during training such that $p_0 = 0.79$ and $p_0 = 0.67$, respectively. The resulting probability distributions are indicated using the diameter of the symbol points. Initialization of the constellation with either 32-APSK or 32-QAM is required to obtain the efficient learned constellations shown in the figure. 
Remaining trainable parameters are initialized with the usual Glorot initialization \cite{glorot2010understanding}. Fig. \ref{gaptocap} evaluates the performance of the learned constellations using BICM channel capacity $\sum_{i=1}^m I(B_i;Y)$ computed using Monte-Carlo simulation. The figure indicates the variation of the SNR gap between the BICM channel capacity and the Gaussian channel capacity with spectral efficiency. In each figure, $P_{X,\theta}$ applicable to trained constellation is used to obtain \lq APSK Shaped'  and \lq QAM Shaped' scenarios. At R = 3 bits/s/Hz, a channel capacity gain exceeding 0.5 dB is visible for both $T_1$ and $T_2$ with probabilistic shaping compared to Uniform 32-APSK or 32-QAM. Note that the BICM channel capacity is closely approximated by $\mathcal{L}(\theta,\theta_D)$ \cite{ait2020joint}.
\begin{figure}[!t]
\centerline{\includegraphics[width=0.8\columnwidth]{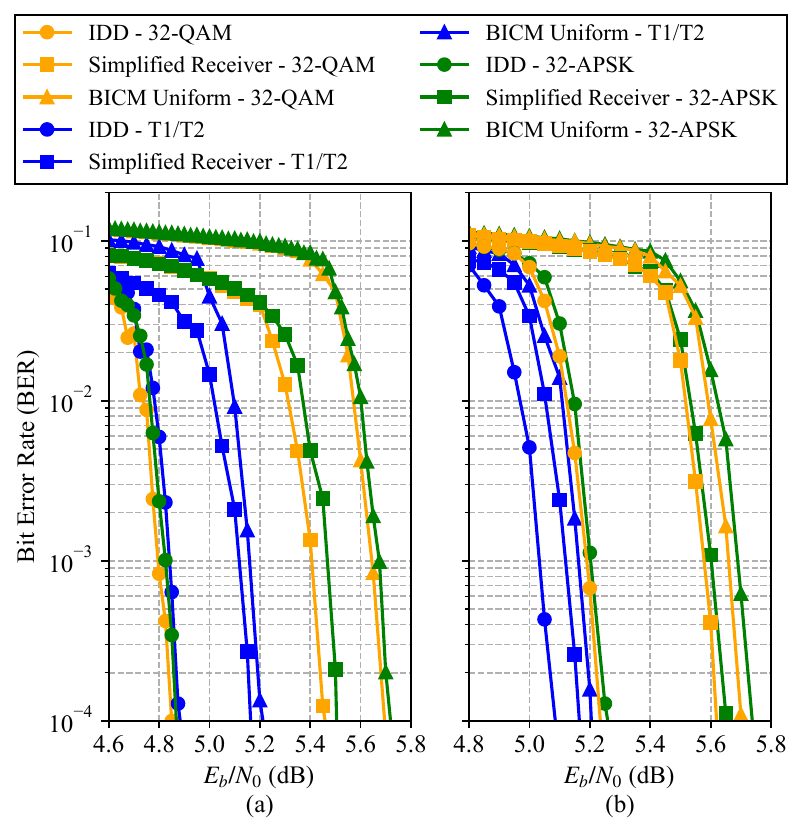}}
\caption{BER performance of the simplified receiver, IDD receiver and the baseline uniform BICM evaluated with 32-APSK, 32-QAM and trained constellations. (a) $S = \{0\}$ with $T_1$ constellation. (b) $S = \{0,4\}$ with $T_2$ constellation.}
\label{berresults}
\end{figure}

To evaluate the BER performance of $T_1$ and $T_2$, a digital video broadcasting satellite second generation (DVB-S2x) \cite{etsi2005digital} LDPC code of $n = 64800$ is used. BP algorithm with maximum 40 iterations is used for FEC decoding. Two shaping encoders are constructed according to the approach detailed in \cite{valenti2012constellation} to closely match the $p_0$ values obtained from training. First shaping encoder is constructed with parameters $k_s=2$, $n_s=4$ and $p_0 = 0.8125$ to implement the $|S| = 1$ scenario. A second shaping encoder with parameters $k_s=3$, $n_s=4$ and $p_0 = 0.6875$ is selected to implement the $|S| = 2$ scenario. A rate-$3/5$ LDPC encoder is used for the uniform system. Using a rate-$2/3$ LDPC encoder with the $R_s$ values corresponding to both cases of $|S| = 1$ and $|S| = 2$, results in an effective rate of $R = 3$ bits/s/Hz. Fig. \ref{berresults}(a) and Fig. \ref{berresults}(b) includes the BER performance of the simplified receiver, IDD receiver and the baseline uniform BICM performance for the cases of $|S| = 1$ and $|S| = 2$, respectively. In $|S| = 1$ scenario, $T_1$ constellation shows a 0.5 dB gain compared to 32-APSK or 32-QAM under uniform BICM. This implies that the geometry of $T_1$ is more efficient compared to other two constellations.  However, all constellations exhibit nearly same performance with the IDD receiver. Note that BER gain in the IDD receiver consists of both the iterative gain and the shaping gain. The simplified scheme with $T_1$ exhibits a gain of 0.3 dB compared to when 32-APSK or 32-QAM is used. In $|S| = 2$ scenario, $T_2$ exhibits better performance in all three schemes. IDD receiver exhibits a BER gain of 0.15 dB with $T_2$, when compared to 32-APSK or 32-QAM. As evident from both the capacity plots in Fig. \ref{gaptocap} and BER plots in Fig. \ref{berresults}, a performance degradation of 32-APSK and 32-QAM is visible when compared to $|S| = 1$. The $T_2$ constellation partially mitigates this performance degradation, improving the performance. 

\subsection{IDD Training Architecture}

\begin{figure}[!t]
\centerline{\includegraphics[width=0.9\columnwidth]{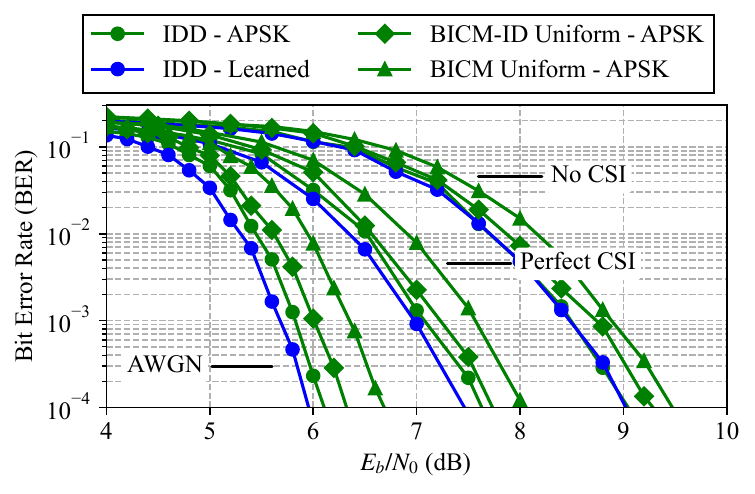}}
\caption{BER performance of the uniform BICM, BICM-ID and shaped IDD system under AWGN and block fading channel conditions.}
\label{berresultsBF}
\end{figure}

To evaluate the performance of the IDD training architecture, a shorter 5G-NR LDPC \cite{3gpp_report} code of length 1440 is used for both training and evaluation. Fig. \ref{berresultsBF} shows the BER performance of the IDD system over AWGN and block fading channels with probabilistically shaped 32-APSK and learned constellations. For comparison, BICM and BICM-ID performance with uniform 32-APSK is included. Performance is evaluated in AWGN, fading with perfect CSI, and fading with no CSI conditions. The block fading channel is characterized by Rician fading with K-factor 10, $N_{p} = 3$, $N_{BF} = 19$ and linear minimum mean square error (LMMSE) channel estimation is performed for no CSI condition. Under fading channels, equalizer output $y^{eq}_j$ and the input symbols $x_j$ are assumed to have an equivalent AWGN channel input-output relationship \cite{dejonghe2002turbo}. Assuming the corresponding channel estimate is given by $\hat{h}_j$ for the received symbol $y_j$, LMMSE equalizer estimate is given by $y^{eq}_j = \frac{y_j \hat{h}_j^*}{|\hat{h}_j|^2 + N_0}$. Equalizer estimate is expressed as $y^{eq}_j = \rho_j x_j + \nu_j$, where $\rho_j = \frac{|\hat{h}_j|^2}{|\hat{h}_j|^2 + N_0}$ and $\nu_j \sim \mathcal{CN}(0, \rho_j^2 - \rho_j)$. An AWGN MAP demapper is used for training under both channel conditions, considering the equivalent AWGN relationship for the block fading channel. Training is performed using Adam \cite{kingma2015adam} optimizer with a learning rate $10^{-3}$, $B = 10^2$, and 5000 training iterations. Within the training batch, AWGN noise samples are generated such that associated $E_b/N_0$ value spread among the waterfall region of the BER curves. The transmitter uses $|S| = 1$ scenario with similar parameters as mentioned in Section \ref{subsec:non_idd}. The IDD training architecture is initialized with $I = 40$ with a single belief propagation iteration per outer iteration \cite{trainable_comm}. For AWGN and perfect CSI conditions, training yielded 0.15dB and 0.1dB gains compared to probabilistically shaped 32-APSK. In the absence of CSI, no such gain is visible, highlighting the susceptibility of the proposed training method for channel estimation errors. Training of IDD system with neural demappers resulted in similar performance as MAP demapper, except with iterative receivers under block fading channel conditions, where sub-optimal BER performance was observed.



\vspace{-2mm}

\section{Conclusion}

This paper has investigated an end-to-end learning-based approach for joint geometric and probabilistic constellation shaping. Subsequently, the method is extended using the deep unfolding technique to incorporate the iterative detection and decoding loop for training. The performance is evaluated with two types of iterative receivers, using two different shaping bit configurations. Under AWGN channel, shaped constellations with the simplified scheme exhibit BER gains reaching 0.3 dB when compared to standard APSK and QAM constellations. Under block fading channel with perfect CSI, the IDD scheme exhibits BER gains of up to 0.1dB with learned constellations.

\bibliographystyle{IEEEtran}
\bibliography{IEEEabrv,references}

\end{document}